\documentclass[onecolumn]{revtex4}
\usepackage{amsmath}
\usepackage{amssymb}
\usepackage{color}
\usepackage{ulem}
\usepackage{graphicx}

\newcommand{\ra}{\rangle}
\newcommand{\la}{\langle}
\newcommand{\pa}{\partial}
\newcommand{\mH}{{\mathfrak{H}}}
\newcommand{\mh}{{\mathfrak{h}}}

\begin{document}

\title{Time Parametrizations in Long-Range Interacting Bose--Einstein Condensates}

\author{Eiji Konishi}
\email{konishi.eiji.27c@kyoto-u.jp}
\address{Graduate School of Human and Environmental Studies, Kyoto University, Kyoto 606-8501, Japan}

\date{\today}

\begin{abstract}
Time-reparametrization invariance in general relativistic space-time does not allow us to single out a time in quantum mechanics in a mechanical way of measurement.
Motivated by this problem, we examine this gauge invariance in the ground state of the quasi-stationary coarse-grained state of a long-range interacting closed system of identical or identified, macroscopic, and spatiotemporally inhomogeneous Bose--Einstein condensates in the thermodynamic and Newtonian limits.
As a result, we find that it is a theoretical counterexample of this gauge invariance, except for proper-time translational invariance, at a coarse-grained level.
\end{abstract}

\maketitle


\section{Introduction}

The {\it problem of time} in canonical quantum gravity\cite{ADM,DeWitt,Regained,Vilenkin,Wald,Kuchar,Isham,Rovelli,Kiefer,Anderson}, namely, the conflict in the concept of time between quantum mechanics and general relativity in the Hamiltonian formalism, is one of the fundamental problems in modern theoretical physics and has been actively debated for decades.
It is well known that, in the quantum context, this problem stems from timelessness in the Wheeler--DeWitt equation
\begin{equation}
\widehat{{\cal H}}|\Psi\ra=0
\end{equation}
for the so-called {\it super-Hamiltonian} operator $\widehat{{\cal H}}$ and the state vector $|\Psi\ra$ of the Universe\cite{Vilenkin,Wald,Kuchar,Isham}.

On the other hand, the semiclassical context where a {\it cousin} of this problem can be approached is quantum field theory in a general relativistic space-time\footnote{The following part of this paragraph is our original formulation of time.
Recently, there have been developments in a distinct formulation of time, that is, the conditional probability interpretation of time\cite{Kuchar,Isham,PW1,PW2,AK}; for some significant examples, see Refs.\cite{QC1,QC2,QC3,QC4,QC5}.}.
For quantum field theory in a general relativistic space-time, we characterize time by two properties:
\begin{enumerate}
\item[(I)] Time is a classical observable {\it $\grave{a}$ la} von Neumann\cite{Neumann} in the exact sense.
\item[(II)] The classical action of general relativity is invariant under reparametrizations of a {\it time parameter} (i.e., the coordinate time of a space-like hypersurface in the $3+1$ decomposition of a space-time\cite{ADM}).
\end{enumerate}
The classicality of time in property (I) means that time eigenstates with different eigenvalues have no mutual quantum interference exactly.
As an illustration of the classicality of time, the Hilbert space $\mh_{N,N^a}$ of the composite of time and the ground state of a condensate as a solitonic solution in quantum field theory has the direct-sum structure
\begin{equation}
\mh_{N,N^a}=\bigoplus_{t\in{\boldsymbol Z}\cdot t_{{\rm pl},t}} \mh_{N,N^a}(t)\;,\label{eq:mH}
\end{equation}
where $t$ is a time parameter playing the role of a superselection charge (i.e., an observable that commutes with all observables), $t_{{\rm pl},t}$ is the Planck time measured by using a time parameter $t$, $\mh_{N,N^a}(t)$ is the one-dimensional superselection sector (i.e., an eigenspace with respect to time) at time $t$, and $N=N(t)$ and $N^a=N^a(t)$ are respectively a dimensionless time-lapse function (i.e., the time-reparametrization gauge) and a shift vector (i.e., the spatial-diffeomorphism gauge) at the quantum mechanical expectation value of the center-of-mass position of the soliton.


Time-reparametrization invariance in the expression (\ref{eq:mH}) of the Hilbert space $\mh_{N,N^a}$ does not allow us to single out a time in quantum mechanics in a {\it mechanical} way of measurement, whereas there is a renowned proposal for singling out a time in a {\it statistical mechanical} way of measurement\cite{Rovelli,Rovelli1,Rovelli2,Rovelli3,Rovelli4,Rovelli5}.
This means that the time parameter $t$ in the Schr$\ddot{{\rm o}}$dinger equation, which is written as
\begin{equation}
i\hbar\frac{\delta |\psi(t)\ra}{\delta t}=N\widehat{H}|\psi(t)\ra\;,\ \ \delta t=t_{{\rm pl},t}\;,
\end{equation}
cannot be singled out at the fundamental level, where the proper time is for the case of $N=1$.
The purpose of this article is to suggest a theoretical counterexample of this gauge invariance, except for proper-time translational invariance, in quantum field theory at a {\it coarse-grained} level and to offer a novel scenario for resolving this problem of {\it singling out a time in quantum mechanics}, in a mechanical way of measurement, without assuming any theoretical hypothesis.

In this article, we choose $(-,+,+,+)$ as the signature of space-time metric tensor $g_{\mu\nu}$ in the square of the line element
\begin{equation}
ds^2=g_{\mu\nu}dx^\mu dx^\nu=-N^2c^2 dt^2+q_{ab}(d\xi^a+ N^a dt)(d\xi^b+N^b dt)\label{eq:ds2}
\end{equation}
in the $3+1$ decomposition of a curved space-time for the Einstein gravity with the time-lapse function $N$, the shift vector $N^a$, and the induced metric tensor $q_{ab}$, take the Newtonian limit (i.e., the non-relativistic and weak-field limits) of the systems to be considered, and adopt the natural system of units in the following parts.
Hatted variables are quantum mechanical operators in the Heisenberg picture unless otherwise noted, and a superscript dot denotes the (total) time derivative.

The rest of this article is organized as follows.
In Sec. II, after conceptual preliminaries, we model the desired system in the thermodynamic and Newtonian limits and describe its kinetic theoretical property.
In Sec. III, we examine time-reparametrization invariance in the Hilbert space of the composite of time and the ground state of this system.
As a result, we find that time-reparametrization invariance, except for proper-time translational invariance, is lost in this Hilbert space at a coarse-grained level.
In Sec. IV, we conclude this article.
In Appendix A, we explain the quasi-particle picture of quantum field theory in the non-relativistic limit and derive two equations presented in Sec. II.


\section{The model}

As conceptual preliminaries for modeling the desired system, we first consider a {\it single} non-relativistic, macroscopic, and spatiotemporally inhomogeneous Bose--Einstein condensate, ${\cal C}$, as a solitonic solution in quantum field theory (e.g., of a scalar field) of a generally covariant Lagrangian density.
Due to the existence of such a spatially inhomogeneous object in the ground state, the global spatial translational symmetry of the system is spontaneously broken.
Then, as the Goldstone modes of this spontaneously broken global symmetry, there exist three {\it quantum mechanical} degrees of freedom of the so-called {\it quantum coordinate}, $\widehat{Q^a}$, and its velocity, $\dot{\widehat{Q^a}}$, which are respectively the center-of-mass position and velocity operators of ${\cal C}$ and are accompanied by no particle mode\cite{Umezawa,Umezawa1,Umezawa2,Umezawa3}.
Here, as a feature of the Goldstone modes of the spontaneously broken spatially translational symmetry, the temporal linearity of $\widehat{Q^a}$ holds as (for its derivation, see Appendix A)
\begin{equation}
\ddot{\widehat{Q^a}}=0\label{eq:ddot}
\end{equation}
for the isolated system of a single condensate ${\cal C}$ in the inertial frame of reference\cite{Umezawa,Umezawa3}.
The role of this quantum coordinate is to rearrange the c-number spatial coordinate $\xi^a$ in quantum field theory in the form of the so-called {\it c--q transmutation} with respect to the spatial translation\cite{Umezawa}
\begin{equation}
\Xi^a_{{\cal C}}=\xi^a-{\cal Q}^a\;,\label{eq:Xi}
\end{equation}
by which the space- and time-dependent {order parameter} (i.e., the vacuum expectation value of the generally non-linear boson Heisenberg field) $\phi_{\cal C}$ of ${\cal C}$ is spatially coordinated as $\phi_{\cal C}=\phi_{\cal C}(\Xi^a_{{\cal C}},t)$\cite{Umezawa,Umezawa1,Umezawa2,Umezawa3}.
By the c--q transmutation rule (\ref{eq:Xi}), the lost global spatial translational symmetry, with respect to a translation $\delta \xi^a$ of c-number $\xi^a$, in the ground state of the system is restored as (i.e., translated into) the symmetry, with respect to the inverse translation $\delta {\cal Q}^a(=-\delta \xi^a)$ of q-number $\widehat{Q^a}$ (denoted by ${\cal Q}^a$), of the quantum mechanical system of the Goldstone modes.
Here, we denote the quantum coordinate $\widehat{Q^a}$ and its non-relativistic velocity $\dot{{\widehat{Q^a}}}$ of a condensate ${\cal C}$, whose quantum fluctuations around their quantum mechanical expectation values are negligible due to macroscopicity of ${\cal C}$, by ${\cal Q}^a$ and $\dot{{\cal Q}}^a$ without hats, respectively.
The degrees of freedom $({\cal Q}^a,\dot{{\cal Q}}^a)$ for such a condensate can be regarded as classical mechanical variables.
Namely, ${\cal C}$ is treated as a classical soliton.


Now, we consider a non-relativistic long-range interacting (specifically, self-gravitating) {closed} system of identical or identified Bose--Einstein condensates $\{{\cal C}\}$ in the thermodynamic limit as the desired system.
Here, the quantum mechanical degrees of freedom $({\cal Q}^a_{{\cal C}},\dot{{\cal Q}}^a_{\cal C})$ of an {\it individual} condensate ${\cal C}$ are not the Goldstone modes, and in the presence of the long-range interactions among the condensates, the temporal linearity (\ref{eq:ddot}) of ${\cal Q}^a_{{\cal C}}$ does not hold in the inertial frame of reference.
Instead of the individual condensates, the Goldstone modes are the three quantum mechanical degrees of freedom of the {\it total} condensate.
However, the c--q transmutation with respect to the spatial translation (\ref{eq:Xi}) holds {in the order parameter} of the boson-transformed Heisenberg field for the space- and time-dependent shift parameter
\begin{equation}
\delta\varphi_{{\cal C}}=\delta\varphi_{{\cal C}}(\xi^a,t;{\cal Q}^a_{\cal C},\dot{{\cal Q}}^a_{\cal C})\;,
\end{equation}
which is not a Heisenberg field but satisfies the free Heisenberg equation in the inertial frame of reference
\begin{equation}
\Lambda(\pa)\delta \varphi_{\cal C}(\xi^a,t;{\cal Q}^a_{\cal C},\dot{{\cal Q}}^a_{\cal C})=0\;,
\end{equation}
of the boson transformation of a free field of quasi-particles $\widehat{\varphi}(\xi^a,t)$ having zero vacuum expectation value and satisfying
\begin{eqnarray}
\Lambda(\pa)\widehat{\varphi}(\xi^a,t)=0
\end{eqnarray}
in the boson Heisenberg field $\widehat{\psi}(\xi^a,t;\widehat{\varphi})$
\begin{eqnarray}
\widehat{\varphi}(\xi^a,t)\to \widehat{\varphi}(\xi^a,t)+\sum_{{\cal C}}\delta\varphi_{{\cal C}}(\xi^a,t;{\cal Q}^a_{\cal C},\dot{{\cal Q}}^a_{\cal C})
\end{eqnarray}
to create each {\it individual} condensate ${\cal C}$ of quasi-particles {\it before} the boson transformation, as
\begin{equation}
\delta\varphi_{\cal C}=\delta\varphi_{\cal C}(\Xi_{\cal C}^a,t)
\end{equation}
under the linear superposition of the boson transformation parameters of all condensates\cite{Umezawa,Umezawa4a,Umezawa4b}
\begin{equation}
\delta\varphi_{\{{\cal C}\}}(\xi^a,t;\{{\cal Q}^a_{\cal C}\},\{\dot{{\cal Q}}^a_{\cal C}\})=\sum_{\cal C}\delta\varphi_{\cal C}(\Xi^a_{\cal C},t)\;,\label{eq:rule}
\end{equation}
where $\xi^a$ in any $\delta\varphi_{{\cal C}}$ can be shifted independently of $\xi^a$ in the other $\delta\varphi_{{\cal C}}$\cite{Umezawa,Umezawa4b}.
Here, the {\it boson transformation} is based on the boson transformation theorem\cite{Umezawa5} (see Appendix A.1) and is a linear way of treating the generally non-linear order parameter in the presence of many condensates $\{{\cal C}\}$\cite{Umezawa}.
The reason for this rule of the c--q transmutation in Eq.(\ref{eq:rule}) is that the generators of the spatial translations of the system $\{{\cal C}\}$ are the quantum mechanical momentum operators of the total condensate (i.e., the sum of the quantum mechanical momentum operators of all condensates) as the Goldstone modes.
Now, in the ground state of the quantum field theoretical sector, which is the vacuum state of quasi-particles {\it after} the boson transformation, the Hamiltonian of this system of the classical solitons $\{{\cal C}\}$ in the quasi-particle picture is (for its derivation, see Appendix A)
\begin{eqnarray}
H_{\{{\cal C}\}}=\sum_{{\cal C}}N_{\cal C}\Biggl\{\frac{{\cal P}_{\cal C}^2}{2M}+\frac{1}{2}\sum_{{\cal C}^{\prime}(\neq {\cal C})}\Phi(|{\cal Q}^a_{\cal C}-{\cal Q}^a_{{\cal C}^\prime}|)\Biggr\}
+\sum_{\cal C}N_{\cal C}^a {\cal P}_{{\cal C},a}\;,\label{eq:Ham}
\end{eqnarray}
where $N_{\cal C}$ and $N_{\cal C}^a$ are respectively the time-lapse function and the shift vector at $\xi^a={\cal Q}^a_{\cal C}$, $M$ denotes the mass of the identical or identified condensate ${\cal C}$,
\begin{eqnarray}
{\cal P}_{\cal C}^a=\frac{M\bigl(\dot{{\cal Q}}^a_{\cal C}-N^a_{\cal C}\bigr)}{N_{\cal C}}=-\frac{M\dot{\Xi}^a_{\cal C}}{N_{\cal C}}\label{eq:Pdef}
\end{eqnarray}
is the canonical momentum (i.e., the center-of-mass momentum) of a condensate ${\cal C}$, and $\Phi$ is the potential energy of a binary long-range interaction, specifically, the Newtonian gravitational interaction.
The third term in Eq.(\ref{eq:Ham}) gives rise to the inertial force.

As a theoretical example of this system, we consider a self-gravitating closed system of hypothetical {\it boson stars}\cite{Bose,ChavanisBook,Bose2} (macroscopic and spatiotemporally inhomogeneous Bose--Einstein condensates bound by self-gravity) of dark matter bosons, to which axions are applicable\cite{Bose2}, in the Universe.
In the phenomenological description, the system of boson stars of non-relativistic bosons with the individual mass $m$ is reduced to the model of the self-interacting Schr${\ddot{{\rm o}}}$dinger field $\psi(x)$ (s.t., $x=(\xi^a,t)$) of bosons with the Lagrangian density in the exactly inertial frame of reference for the {Newtonian gravity}\cite{ChavanisBook}
\begin{equation}
{\cal L}=i\psi^\dagger(x) {\dot{\psi}(x)}-\frac{1}{2m}|\nabla \psi(x)|^2-m U(x)|\psi(x)|^2-\frac{g}{2}|\psi(x)|^4-\frac{1}{8\pi} |\nabla U(x)|^2\label{eq:BoseLag}
\end{equation}
for the coupling constant $g$ of the short-range repulsive/attractive interaction, subject to the Newtonian gravitational potential $U(x)$, which is an auxiliary field.
In the boson stars as solitonic condensates, the quantum pressure arising from the Heisenberg uncertainty principle and the pressure arising from the short-range interaction are in balance with the attractive gravitational interaction\cite{ChavanisBook}.
The Lagrangian density (\ref{eq:BoseLag}), which is defined for the exactly inertial frame of reference in the flat space-time
\begin{equation}
ds^2=-dt^2+\delta_{ab}d \xi^ad \xi^b\;,\label{eq:Exact}
\end{equation}
is the equivalently replaced (with respect to gravity, owing to the Newtonian limit) and reduced form of the generally covariant Lagrangian density of the Einstein gravity and the boson field coupled to gravity, which is defined for the space-time metric tensor $g_{\mu\nu}$ in the curved space-time
\begin{equation}
ds^2=g_{\mu\nu}dx^\mu dx^\nu=-(1+2U(x))dt^2+(1-2U(x))\delta_{ab}d \xi^a d \xi^b\;.\label{eq:dsU}
\end{equation}
Equation (\ref{eq:dsU}) gives an approximately inertial frame of reference if $U(x)\ll 1$ holds.


Next, as a kinetic theoretical property of a generic non-relativistic long-range interacting classical system of identical particles (assuming identical or identified condensates $\{{\cal C}\}$) in the presence of collective effects (we assume collective effects in the system $\{{\cal C}\}$: $\{{\cal C}\}$ starts from a dynamically unstable initial distribution)\cite{PL2,Book,PR}, this system undergoes collisionless phase mixing of particles along their own constant-energy hypersurfaces in the $\mu$-space, governed by the Vlasov kinetic equation (i.e., the collisionless Boltzmann equation), under spatiotemporally strong initial oscillations of the mean-field potential energy of the long-range interaction, which significantly promote phase mixing.
Collisional relaxation (i.e., thermodynamical relaxation) of the system occurs as a result of the finite particle-number effect (i.e., graininess of the system; in three spatial dimensions, as a result of the $n$-body correlations with $n \ge 2$, which are slower than phase mixing), and it was proved by Braun and Hepp that, in the thermodynamic limit of the particle number of the system, the lifetime of the collisionless regime before the collisional regime diverges\cite{Book,PR,Braun}.
After time elapses on the dynamical time scale, the coarse-grained distribution of the system converges to the so-called {\it quasi-stationary state} (QSS), which is the collisionless equilibrium state of the system, as a result of phase mixing\cite{PL2,Book,PR}.

\begin{figure}[htbp]
\begin{center}
\includegraphics[width=0.4 \hsize,bb=4 5 258 259]{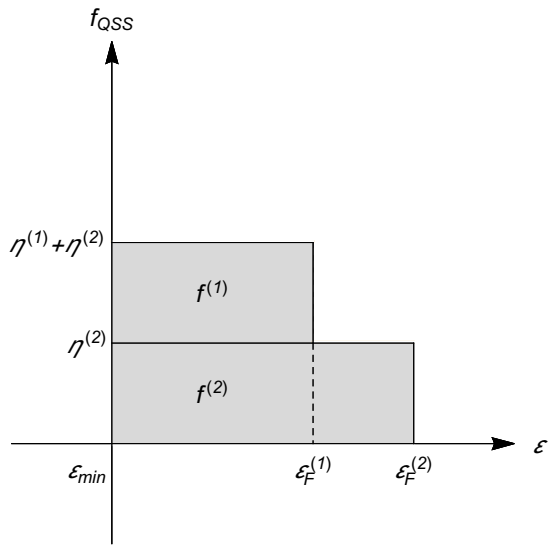}
\end{center}
\caption{The inferred zero-temperature QSS distribution (\ref{eq:QSS}) of the system $\{{\cal C}\}$ from the results of Refs.\cite{KS,KS2}.
This ground-state distribution is a two-step functional of the one-condensate energy function $\varepsilon=\varepsilon({\cal Q}^a,{{{\cal P}}^a})$.
The most remarkable property of $f_{\rm QSS}$ is that there {\it coexist} two different Fermi energies: $\varepsilon_F^{(1)}$ and $\varepsilon_F^{(2)}$.}
\end{figure}

From the results of Refs.\cite{KS,KS2}, we infer that, in the $\mu$-space, the coarse-grained distribution of this system in the QSS, in the presence of collective effects, at zero temperature is the {\it superposition} of two energy step functions (see Fig. 1)
\begin{eqnarray}
f_{\rm QSS}=\sum_{i=1}^2f^{(i)}=\sum_{i=1}^2 \eta^{(i)}\theta(\varepsilon_F^{(i)}-\varepsilon)\label{eq:QSS}
\end{eqnarray}
for the one-particle energy function $\varepsilon={\cal P}^2/2M+\Phi_{\rm MF}({\cal Q}^a)$ with the mean-field potential energy $\Phi_{\rm MF}({\cal Q}^a)$ (i.e., the integral of the binary potential energy $\Phi(|{\cal Q}^a-{\cal Q}^{a\prime}|)$ multiplied by the coarse-grained distribution, $f({\cal Q}^{a\prime},{{{\cal P}}^{a\prime}},t)$, of the system over the $\mu$-space), the so-called {\it Fermi energy} $\varepsilon_F$\footnote{This term comes from the Fermi energy of the Fermi--Dirac distribution at zero temperature.}, the diluted phase-space density $\eta$, and the Heaviside step function $\theta$.
Here, each component $f^{(i)}$ $(i=1,2)$ as a subsystem is the zero-temperature case of the Lynden--Bell ergodic distribution\cite{LB}, which is the maximum entropy distribution in the coarse-grained $\mu$-space subject to the total particle number, total energy conservation laws, and the incompressibility of the diluted phase-space density $\eta^{(i)}$\cite{KS,KS2}, where the uniqueness of the non-zero fine-grained phase-space density $\eta_0=\eta^{(1)}+\eta^{(2)}$ in the system $\{{\cal C}\}$ is assumed, in the collisionless regime\footnote{In Ref.\cite{KS}, the superposition of two Lynden--Bell distributions was discovered as the QSS in the simplest long-range interacting classical system as the solution of the core-halo problem in this system.
The core-halo problem and its ansatz were proposed in Ref.\cite{PL1} and investigated in Ref.\cite{PL2}.}.
This incompressibility, which follows from the equations $\dot{f}^{(i)}=0$ $(i=1,2)$ of the dynamical relaxation between the two subsystems at zero temperature (i.e., in their fine-grained distributions with diluted phase-space densities) in the same form as the Vlasov kinetic equation\cite{KS,KS2}, plays the same role as the Pauli exclusion principle for fermions in the statistics.
This fact is the origin of similarity between the Lynden--Bell distribution in classical statistics and the Fermi--Dirac distribution.


Finally, the motivation for considering system (\ref{eq:Ham}) in our problem is the c--q transmutation (\ref{eq:Xi}).
A consequence of the c--q transmutation (\ref{eq:Xi}) is translation of the velocity $\dot{{\cal Q}}^a_{\cal C}$ of a condensate ${\cal C}$ into the inverse velocity, $N_{{\rm phys},{\cal C}}^a$, of the c-number spatial coordinate $\xi^a$ in the boson transformation parameter $\delta\varphi_{\cal C}$ at $\xi^a={\cal Q}_{\cal C}^a$.
Namely, in the c--q transmutation (\ref{eq:Xi}), we equate
\begin{equation}
-\dot{{\cal Q}}^a_{{\cal C}}=N^a_{{\rm phys},{\cal C}}(\xi^a={\cal Q}^a_{\cal C})\label{eq:main}
\end{equation}
for each ${\cal C}$ in $\{{\cal C}\}$, with $({\cal Q}^a_{\cal C},\dot{{\cal Q}}^a_{\cal C})$.
Here, we make three remarks:
\begin{enumerate}
\item[(i)] Because $N^a_{{\rm phys},{\cal C}}$ is spatial-diffeomorphism gauge independent, $N^a_{{\rm phys},{\cal C}}$ is a time-reparametrization gauge as well as the time-lapse function $N$.

\item[(ii)] Both sides of Eq.(\ref{eq:main}) have no field dimensions but do have particle dimensions [L$\cdot$T$^{-1}$].

\item[(iii)] The quantum mechanical degrees of freedom of the total condensate are the zero-energy modes of the quasi-particle field in the boson-transformed Heisenberg field\cite{Umezawa}, so they are not dropped off in the ground state of the system $\{{\cal C}\}$.
\end{enumerate}


\section{Time parametrizations}

In this section, we examine time-reparametrization invariance in the Hilbert space of the composite of time and the ground state of the coarse-grained system $\{{\cal C}\}$.

Because we treat quantum coordinate ${\cal Q}^a$ and its canonical conjugate ${\cal P}^a$ as classical variables, by combining Eq.(\ref{eq:Xi}) with Eq.(\ref{eq:QSS}) we find that the Hilbert space $\mH_{\{{\cal C}\},N,N^a}$ of the composite of time and the ground state of the coarse-grained system $\{{\cal C}\}$ with a fine-grained time-lapse function $N=N(x)$ and a fine-grained shift vector $N^a=N^a(x)$ decomposes as
\begin{eqnarray}
\mH_{\{{\cal C}\},N,N^a}&=&\mH^{(1)}_{N,N^a}\otimes \mH^{(2)}_{N,N^a}\label{eq:mH12}\\
&=&\Biggl(\bigotimes_{({\cal Q}^a,{{{\cal P}}^a})\in f^{(1)}}\bigoplus_1^{f^{(1)}({\cal Q}^a,{{{\cal P}}^a})\omega}\mh_{({\cal Q}^a,{{{\cal P}}^a}),N,N^a}\Biggr)
\otimes \Biggl(\bigotimes_{({\cal Q}^a,{{{\cal P}}^a})\in f^{(2)}}\bigoplus_1^{f^{(2)}({\cal Q}^a,{{{\cal P}}^a})\omega}\mh_{({\cal Q}^a,{{{\cal P}}^a}),N,N^a}\Biggr)\;,\label{eq:mH3}
\end{eqnarray}
where, in the occupied bins in the $\mu$-space, as the equivalence between gauges with respect to time reparametrizations, a fine-grained time-lapse function $N(x)$ (i.e., the gauge of a time parametrization $t$) at an arbitrary fine-grained position $\xi^a={\cal Q}^a$ (i.e., at $\Xi^a=0$) is equivalent to each fine-grained pair $({\cal Q}^a,\dot{{\cal Q}}^a)$ in a set, fully specified by ${\cal Q}^a$, by the c--q transmutation rule (\ref{eq:Xi}); note the previous remark (i).
Here, bin widths of ${\cal Q}^a$ and ${\cal P}^a$ are contravariant and their $\mu$-space volume is denoted by $\omega$.
We formally set $\omega$ as $1/\eta_0$.

\begin{figure}[htbp]
\begin{center}
\includegraphics[width=0.4 \hsize,bb=4 5 259 256]{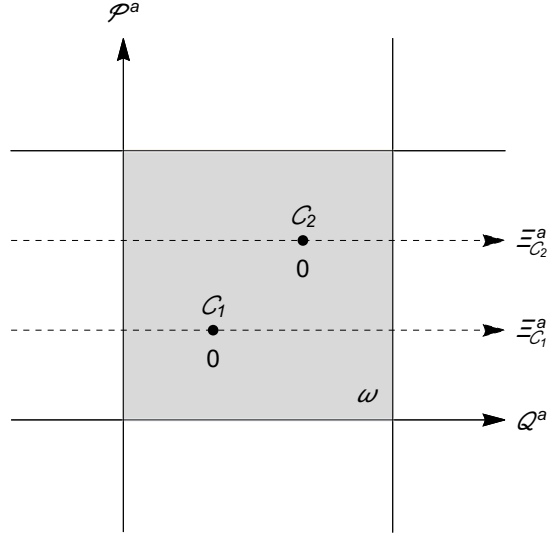}
\end{center}
\caption{A bin having volume $\omega$ in the $\mu$-space (the gray domain), which is occupied by two condensates ${\cal C}_1$ and ${\cal C}_2$ in one of the Lynden--Bell subsystems.
The rearranged spatial coordinate systems $(\Xi^a_{{\cal C}_1},\dot{\Xi}^a_{{\cal C}_1})$ and $(\Xi^a_{{\cal C}_2},\dot{\Xi}^a_{{\cal C}_2})$, which are associated with ${\cal C}_1$ and ${\cal C}_2$, respectively, are originally {\it distinguishable} (i.e., inequivalent under any spatial diffeomorphism) but are {\it identical} after the coarse-graining of the system distribution.
After this coarse-graining, a fine-grained time-lapse function $N(x)$ and a fine-grained shift vector $N^a(x)$, which are associated with a condensate, are defined over all fine-grained positions in the bin that this condensate occupies.
This is because these fine-grained positions correspond to the coarse-grained position of the bin from the definition of the latter.}
\end{figure}

Equation (\ref{eq:mH3}) is our main result and is derived by noting two facts:
\begin{enumerate}
\item[(A)] The composite of the canonical degrees of freedom $\{({\cal Q}^a,{{{\cal P}}^a})\}$ of condensates is always multipartite (i.e., the tensor product).

\item[(B)] The composite of the originally {\it distinguishable} (i.e., inequivalent under any spatial diffeomorphism\footnote{This is because the fine-grained Lagrangian degrees of freedom $({\cal Q}^a,{{\dot{{\cal Q}}}^a})$ of a condensate ${\cal C}$ in definition (\ref{eq:Xi}) of the rearranged spatial coordinate system $(\Xi^a_{\cal C},\dot{\Xi}^a_{\cal C})$ are independent from the spatial diffeomorphism.}) $\nu$ number of rearranged spatial coordinate systems
\begin{equation}
(\Xi^a_{{\cal C}_1},\dot{\Xi}^a_{{\cal C}_1})\;,\ \ldots\;,\ (\Xi^a_{{\cal C}_{\nu}},\dot{\Xi}^a_{{\cal C}_{\nu}})\label{eq:copies}
\end{equation}
with an {\it identical} coarse-grained pair $(\Xi^a,\dot{\Xi}^a/N)$ in an occupied bin $({\cal Q}^a,{{{\cal P}}^a})$ of $f^{(i)}$ ($i=1,2$) is unipartite (i.e., the direct sum of ${\nu}=f^{(i)}({\cal Q}^a,{\cal P}^a)\omega$ number of copies (\ref{eq:copies}) after the coarse-graining of the system distribution).
Here, the boson transformation parameter $\delta\varphi_{{\cal C}}(\Xi^a_{\cal C},t)$ of a condensate ${\cal C}$ at
\begin{equation}
\Xi^a_{{\cal C}_k}=0\;,\ \ {\cal P}^a_{\cal C}=-\frac{M\dot{\Xi}^a_{{\cal C}_k}}{N_{{\cal C}_k}}\;,\ \ k\in \{1,\ldots,{\nu}\}
\end{equation}
is spatially coordinated by each of Eq.(\ref{eq:copies}).
Note Eq.(\ref{eq:Pdef}) for a given $N_{\cal C}$ and see Fig. 2.
\end{enumerate}

Now, further decomposition of the Hilbert space $\mH_{\{{\cal C}\},N,N^a}$ differs substantially from Eq.(\ref{eq:mH}):
\begin{eqnarray}
\mH_{\{{\cal C}\},N,N^a}=\Biggl(\bigoplus_{\{t_1\}} {\mH}^{(1)}_{N,N^a}(\{t_1\})\Biggr)\otimes\Biggl(\bigoplus_{\{t_2\}} {\mH}^{(2)}_{N,N^a}(\{t_2\})\Biggr)\label{eq:mH2}
\end{eqnarray}
for two types of superselection sectors ($i=1,2$)
\begin{equation}
\mH^{(i)}_{N,N^a}(\{t_i\})\equiv \bigotimes_{({\cal Q}^a,{{{\cal P}}^a})\in f^{(i)}}\mH^{(i)}_{({\cal Q}^a,{{{\cal P}}^a}),N,N^a}(t_i)\;,\label{eq:mHN0}
\end{equation}
where each $t_i$ in $\{t_i\}$ is a ``{\it diluted}'' time parameter (not a reparametrized time parameter) of $t$ and corresponds to an occupied bin $({\cal Q}^a,{{{\cal P}}^a})$ of $f^{(i)}$, and
\begin{eqnarray}
\bigoplus_{t_i\ {\rm over}\ t_{{\rm pl},t}}\mH^{(i)}_{({\cal Q}^a,{{{\cal P}}^a}),N,N^a}(t_i)\simeq \bigoplus^{f^{(i)}({\cal Q}^a,{{{\cal P}}^a})\omega}_1{\mh}_{({\cal Q}^a,{{{\cal P}}^a}),N,N^a}(t)\label{eq:mHN}
\end{eqnarray}
holds, in contrast to the conventional
\begin{equation}
\mH^{(i)}_{N,N^a}(t)=\bigotimes_{({\cal Q}^a,{{{\cal P}}^a})\in f^{(i)}}{\mh}_{({\cal Q}^a,{{{\cal P}}^a}),N,N^a}(t)\;,
\end{equation}
which arises when each element ${\cal C}$ of the system $\{{\cal C}\}$ is {\it not} a spatiotemporally inhomogeneous Bose--Einstein condensate: the c--q transmutation rule (\ref{eq:Xi}) does not hold.
From Eq.(\ref{eq:mHN}), it follows that $t_{{\rm pl},\tau}=Nt_{{\rm pl},t}$ for the proper time $\tau$ is equal to
\begin{eqnarray}
\bigl(f^{(1)}({\cal Q}^a,{{{\cal P}}^a})\omega N\bigr)t_{{\rm pl},t_1}=\bigl(f^{(2)}({\cal Q}^a,{{{\cal P}}^a})\omega N\bigr)t_{{\rm pl},t_2}\label{eq:common}
\end{eqnarray}
at each common occupied bin $({\cal Q}^a,{{{\cal P}}^a})$ of $f^{(1)}$ and $f^{(2)}$ because the superselection sectors are one-dimensional.
Then, this Hilbert space $\mH_{\{{\cal C}\},N,N^a}$ is not invariant under time reparametrizations of $t$, except for proper-time translations\footnote{This is because the time-lapse function does not change under proper-time translations.}, due to the incompatibility between $f^{(1)}$ and $f^{(2)}$.
Namely, two distinct subsystems $f^{(1)}$ and $f^{(2)}$ of $f_{\rm QSS}$ give rise to the different ``{\it diluted}'' time parameters of their superselection sectors (\ref{eq:mHN0}) at each common occupied bin of $f^{(1)}$ and $f^{(2)}$, and then $f^{(1)}$ and $f^{(2)}$ contradict each other at this bin: the discrepancy between the ``{\it diluted}'' time-lapse functions of $t_1$ and $t_2$
\begin{equation}
f^{(1)}({\cal Q}^a,{{{\cal P}}^a})\omega N \not\equiv f^{(2)}({\cal Q}^a,{{{\cal P}}^a})\omega N\label{eq:disc}
\end{equation}
holds in Eq.(\ref{eq:common}) for the fine-grained time-lapse function $N(x)$ at an arbitrary fine-grained position $\xi^a={\cal Q}^a$ in the bin.
The crucial point here is that, in the $\mu$-space region of overlap between the supports of $f^{(1)}$ and $f^{(2)}$, it is, in principle, impossible to specify the subsystem $f^{(i)}$ ($i=1,2$) of $f_{\rm QSS}$ to which each condensate belongs: this is the most remarkable property of $f_{\rm QSS}$ (see Fig. 1).
Because of this point, in this $\mu$-space region, a time reparametrization must be applied for both subsystems $f^{(1)}$ and $f^{(2)}$ of $f_{\rm QSS}$ by using a single time-lapse function (i.e., the gauge of a single time parametrization).
However, for the discrepancy (\ref{eq:disc}), the invariance of the Hilbert space $\mH_{\{{\cal C}\},N,N^a}$ under time reparametrizations of $t$, except for proper-time translations, requires two distinct time-lapse functions.


Here, we add two remarks:
\begin{enumerate}
\item[(iv)] In the thermodynamic limit of the system $\{{\cal C}\}$, we can define the statistical mechanical equilibrium states (i.e., the coarse-grained distributions at equilibria) of $\{{\cal C}\}$ with negligible fluctuations in global quantities\cite{Book}.
The reason why we consider the ground state of $\{{\cal C}\}$ as the statistical mechanical equilibrium state is that the quantum mechanical degrees of freedom of the total condensate are the zero-energy modes of the quasi-particle field in the boson-transformed Heisenberg field, and quantum field theory is defined at zero temperature (cf., remark (iii)).
Here, recall that the collisionless equilibration of $\{{\cal C}\}$ is driven by the potential energy $\Phi$, which is independent of quasi-particles, and the ground state is the vacuum state of quasi-particles.
It is noteworthy that, in the thermodynamic limit, the ground state of $\{{\cal C}\}$ is a classical pure state (i.e., the probability of finding the system in this classical state is unity) with respect to each Lynden--Bell subsystem.

\item[(v)] In our description of the dynamics of the system $\{{\cal C}\}$ in the thermodynamic limit, there implicitly exist coarse-graining scales of ${\cal Q}^a$ and ${{{\cal P}}^a}$ (i.e., small but finite bin widths with their volume $\omega$ in the $\mu$-space)\cite{Tolman} which satisfy $\omega=1/\eta_0$ and the criterion that their coarse-grained distribution converges to the QSS distribution as a stable stationary solution of the Vlasov kinetic equation (cf., the fine-grained distribution of a long-range interacting system never converges to the QSS distribution)\cite{Chavanis}.
\end{enumerate}

\section{Conclusion}

We now conclude this article.
Our study of the QSS of the coarse-grained system $\{{\cal C}\}$ at zero temperature in the thermodynamic and Newtonian limits shows that a long-range interacting (specifically, self-gravitating) closed system of identical or identified, macroscopic, and spatiotemporally inhomogeneous Bose--Einstein condensates is a theoretical counterexample of time-reparametrization invariance, except for proper-time translational invariance, in the ground state of a {\it coarse-grained} quantum field system in a general relativistic space-time, which is replaced with the $3+1$ decomposition of the flat space-time for the Newtonian gravity owing to the Newtonian limit.
(Of course, the classical action of the system is time-reparametrization invariant.)
Namely, in the ground state of this system (cf., remark (iv)), the time parametrizations are mutually inequivalent except for proper-time translational equivalence.
This mutual inequivalence of the equivalence classes of time parametrizations holds over the spatial support, $V_{\{{\cal C}\}}$, of both of the Lynden--Bell distributions $f^{(1)}$ and $f^{(2)}$ of the system $\{{\cal C}\}$ at zero temperature in Eq.(\ref{eq:QSS}), because the fine-grained time-lapse function $N(x)$ and the fine-grained shift vector $N^a(x)$ that appear in both of the Hilbert spaces $\mH^{(1)}_{N,N^a}$ and $\mH^{(2)}_{N,N^a}$ in Eq.(\ref{eq:mH12}) are defined over $V_{\{{\cal C}\}}$.
In general, an equivalence class of time parametrizations accompanies readout of the motion of a mechanical object (i.e., an object in a pure state) within $V_{\{{\cal C}\}}$ by a measuring system.
In our case, as opposed to the conventional case, since the time parametrizations are mutually inequivalent except for proper-time translational equivalence, this means an equivalence class of time parametrizations with respect to proper-time translational equivalence (i.e., a time in quantum mechanics) is singled out.
The proposal of this novel scenario to resolve our problem, in a mechanical way of measurement, concludes this article.

\begin{appendix}
\section{Quasi-particle picture}

In this appendix, we explain the quasi-particle picture of quantum field theory\cite{Umezawa,UmezawaChap,UmezawaApp} in the non-relativistic limit.
Some parts of this appendix accord with the ideas in Ref.\cite{UmezawaApp}.
Because all variables except for those in Eq.(\ref{eq:B0}) and the equations from Eq.(\ref{eq:grav0}) to Eq.(\ref{eq:potfin}) are quantum mechanical operators, we omit their hats.

\subsection{Weak relations}

We suppose a generic original non-relativistic Galilean covariant Lagrangian density
\begin{equation}
{\cal L}={\cal L}[\psi(x),\dot{\psi}(x)]\label{eq:B0}
\end{equation}
of a classical scalar field $\psi=\psi(x)$ (s.t., $x=(\xi^a,t)$) in the absence or presence of an auxiliary field.
Then, from Eq.(\ref{eq:B0}), we obtain the Heisenberg equation of the boson Heisenberg field $\psi(x)$
\begin{equation}
\Lambda(\pa)\psi(x)=F[\psi(x)]\label{eq:B1}
\end{equation}
for a functional $F$, subject to the canonical commutation relation
\begin{equation}
[\psi(x),\pi_{\psi}(y)]\delta(t_x-t_y)=i\delta(x-y)\label{eq:B15}
\end{equation}
for the canonical conjugate $\pi_{\psi}$ of $\psi$.
By simultaneously finding the extended Fock space of quasi-particles, with their annihilation operators $\alpha_k$ for modes $k$, of a ``{\it free}'' field $\varphi=\varphi(x)$ and zero-energy modes and finding a weak relation for the boson Heisenberg field
\begin{eqnarray}
\psi(x)&\overset{{\rm w}}{=}&\psi(x;\vec{q},\vec{p},\varphi)\\
&=&\psi(x;\vec{q},\vec{p},\alpha_k,\alpha_k^\dagger)
\end{eqnarray}
among the matrix elements in this extended Fock space of quasi-particles and zero-energy modes (the so-called {\it dynamical map} of $\psi$), consistent with Eqs.(\ref{eq:B1}) and (\ref{eq:B15}), we solve the Heisenberg equation (\ref{eq:B1}) in a self-consistent way\footnote{In this appendix, we denote the quantum coordinate and its canonical conjugate of a single condensate in the Schr$\ddot{{\rm o}}$dinger picture\cite{Umezawa3,UmezawaApp} by $\vec{q}$ and $\vec{p}$, respectively.
In this picture of $\vec{q}$ and $\vec{p}$, the extended Fock space is the {\it direct product} of the Hilbert space of the quantum mechanical sector and the Fock space of the quantum field theoretical sector\cite{Umezawa3}.}.
Here, a {\it weak relation}
\begin{equation}
X\overset{{\rm w}}{=}Y
\end{equation}
between two operators $X$ and $Y$ means equality between their matrix elements
\begin{equation}
\la a|X|b\ra=\la a|Y|b\ra
\end{equation}
for both of vectors $|a\ra$ and $|b\ra$ in an extended Fock space\cite{Umezawa}.

Note that, for a space- and time-dependent c-number shift parameter $\delta \varphi=\delta \varphi(x)$ satisfying the same equation as $\varphi$,
\begin{equation}
\psi^{\delta \varphi}(x)\overset{{\rm w}}{=}\psi(x;\vec{q},\vec{p},\varphi+\delta\varphi)
\end{equation}
is the dynamical map of the boson-transformed Heisenberg field $\psi^{\delta \varphi}$, which satisfies the Heisenberg equation (\ref{eq:B1}).
This is the original content of the boson transformation theorem\cite{Umezawa,Umezawa5}.

Now, when we solve the Heisenberg equation (\ref{eq:B1}), we obtain the weak relations for the Hamiltonian $H=H(\psi)$ and the momentum $P_a=P_a(\psi)$ of a single condensate
\begin{eqnarray}
H&\overset{{\rm w}}{=}&H(\vec{p},\alpha_k,\alpha_k^\dagger)\;,\\
P_a&\overset{{\rm w}}{=}&P_a(\vec{p},\alpha_k,\alpha_k^\dagger)
\end{eqnarray}
in the extended Fock space of the quasi-particles and the zero-energy modes.
Here,
\begin{equation}
\vec{P}\overset{{\rm w}}{=}\vec{p}\label{eq:Pp}
\end{equation}
follows by assuming asymptotic conditions on quasi-particles\cite{Umezawa,UmezawaApp}.
It is noteworthy that, in the infinite volume limit of the system, $H$ is a function of only $\vec{p}$ and the number operators $n_k=\alpha_k^\dagger \alpha_k$ of quasi-particles: in this limit of the system\cite{Umezawa}, quasi-particle interaction terms, which are always inversely proportional to the system size, drop from $H$ but contribute to the quasi-particle self-energy, which renormalizes the free Hamiltonian of quasi-particles, and the S-matrix, which requires an infinite time process.
In the next subsection, we will determine $H$.

\subsection{Derivations of Eqs.(\ref{eq:ddot}) and (\ref{eq:Ham})}

In this subsection, we derive Eqs.(\ref{eq:ddot}) and (\ref{eq:Ham}) in Sec. II.

First, according with the ideas in Ref.\cite{UmezawaApp}, we determine $H$ in the quasi-particle picture.

In the Schr$\ddot{{\rm o}}$dinger picture, we define
\begin{equation}
{Q}_a={C}_a{M}^{-1}\label{eq:A1}
\end{equation}
for a single condensate.
Here, ${C}_a$ are the Galilean boost operators, and $M$ is the mass operator, which is a Casimir element in the Galilean algebra.
From the Galilean algebra, ${C}_a$ satisfy the commutation relations with $P_b$
\begin{equation}
[{C}_a,{P}_b]=i\delta_{ab}{M}\label{eq:A2}
\end{equation}
and the commutation relations with $H$
\begin{equation}
[{C}_a,{H}]=i{P}_a\;.\label{eq:A3}
\end{equation}
From Eqs.(\ref{eq:A1}) and (\ref{eq:A2}), we obtain the canonical commutation relations
\begin{equation}
[{Q}_a,{P}_b]=i\delta_{ab}\;.\label{eq:A4}
\end{equation}

Due to Eqs.(\ref{eq:Pp}) and (\ref{eq:A4}), $Q_a$ decomposes into
\begin{equation}
{Q}_a={q}_a+\delta {Q}_a(\vec{p},{\alpha}_k,{\alpha}^\dagger_k)\;.\label{eq:A5}
\end{equation}
From Eqs.(\ref{eq:A1}) and (\ref{eq:A3}), we obtain
\begin{equation}
[{Q}_a,{H}]=i{p}_a{M}^{-1}\;.\label{eq:A6}
\end{equation}
Then, from Eqs.(\ref{eq:A5}) and (\ref{eq:A6}), we obtain
\begin{equation}
[\delta{Q}_a,{H}]=i\biggl({p}_a{M}^{-1}-\frac{\pa }{\pa p_a}{H}\biggr)\;.\label{eq:A7}
\end{equation}
In order for Eq.(\ref{eq:A7}) to hold, $\delta Q_a$ must be a function of $\vec{p}$ and $n_k$ because $M$ is a function of $n_k$, and $H$ is a function of $\vec{p}$ and $n_k$.
Thus, $\delta Q_a$ and $H$ commute with each other.
From this fact and Eq.(\ref{eq:A7}), we obtain
\begin{equation}
\frac{\pa}{\pa p_a}{H}(\vec{p},n_k)={p}_a{M}^{-1}\;.\label{eq:A8}
\end{equation}
By solving Eq.(\ref{eq:A8}), we obtain the Hamiltonian in the presence of a single condensate as
\begin{equation}
H(\vec{p},n_k)=M(n_k)+\frac{p^2}{2}M^{-1}(n_k)\;,\label{eq:BH}
\end{equation}
where the first term is the rest energy for the mass operator
\begin{equation}
M(n_k)=M_0+H_0(n_k)
\end{equation}
with the c-number mass $M_0$ of the condensate and the renormalized free Hamiltonian $H_0(n_k)$ of quasi-particles\cite{UmezawaApp}, in the inertial frame of reference.

When the condensate is absent, $\vec{p}=\vec{0}$ and $M_0=0$ hold.
Then, in the quasi-particle picture,
\begin{equation}
H\overset{{\rm w}}{=}H_0(n_k)\label{eq:BH1}
\end{equation}
follows from Eq.(\ref{eq:BH}).
This is the same result as mentioned in the last subsection\cite{Umezawa}.

Next, we derive Eq.(\ref{eq:ddot}).
We define
\begin{eqnarray}
Q_a(t)&=&e^{iHt}Q_ae^{-iHt}\;,\\
q_a(t)&=&e^{iHt}q_ae^{-iHt}\;.
\end{eqnarray}
Then, from Eqs. (\ref{eq:A5}) and (\ref{eq:A6}), we obtain
\begin{eqnarray}
\dot{Q}_a(t)&=&i[H,Q_a(t)]=p_aM^{-1}\;,\\
\dot{q}_a(t)&=&\dot{Q}_a(t)\;.
\end{eqnarray}
From these, we obtain
\begin{equation}
\ddot{Q}_a(t)=\ddot{q}_a(t)=0
\end{equation}
in the inertial frame of reference.
$\ddot{q}_a(t)=0$ is nothing but Eq.(\ref{eq:ddot}).

Finally, we derive Eq.(\ref{eq:Ham}).
We consider a non-relativistic self-gravitating closed system of identical or identified macroscopic condensates (e.g., boson stars) $\{{\cal C}\}$ in the vacuum state of quasi-particles (i.e., $M=M_0$) and in the weak-field limit.
Because of the result (\ref{eq:BH}) for a single condensate in the inertial frame of reference, the Lagrangian of the system subject to {Newtonian gravity} is given in the exactly inertial frame of reference by
\begin{equation}
L_{\{{\cal C}\}}=M\sum_{\cal C} \frac{|\dot{{q}}_{\cal C}^a|^2}{2}-M\sum_{\cal C}U({q}_{\cal C}^a,t)-\frac{1}{8\pi}\int d^3\xi |\nabla U(x)|^2\;,\label{eq:grav0}
\end{equation}
where $U(x)$ is the Newtonian gravitational potential and is an auxiliary field.
The equation of motion of $U(x)$ is the Poisson equation
\begin{equation}
\Delta U(x)=4\pi M \sum_{\cal C}\delta(\xi^a-q^a_{\cal C})\label{eq:Poisson}
\end{equation}
for the Laplacian $\Delta$.
By solving Eq.(\ref{eq:Poisson}), we obtain
\begin{equation}
U(x)=-\sum_{\cal C}\frac{M}{|\xi^a-q^a_{\cal C}|}\;.\label{eq:grav2}
\end{equation}
By substituting Eq.(\ref{eq:grav2}) into Eq.(\ref{eq:grav0}) and using Eq.(\ref{eq:Poisson}) in the integral by parts, the Lagrangian of the system in the exactly inertial frame of reference becomes
\begin{equation}
{L}_{\{{\cal C}\}}=M\sum_{\cal C}\frac{|\dot{{q}}_{\cal C}^a|^2}{2}+\frac{1}{2}\sum_{{\cal C}\neq {\cal C}^\prime}\frac{M^2}{|{q}_{\cal C}^a- {q}^a_{{\cal C}^\prime}|}\;.\label{eq:Lag2}
\end{equation}
In the $3+1$ decomposition of the flat space-time for the Newtonian gravity
\begin{equation}
ds^2=g_{\mu\nu} dx^\mu dx^\nu=-N^2dt^2+\delta_{ab}(d\xi^a+N^a dt)(d\xi^b+N^b dt)\;,\label{eq:31}
\end{equation}
we have the Lagrangian of the system
\begin{eqnarray}
L_{\{{\cal C}\}}&=&\sum_{\cal C}\sqrt{-g(q_{\cal C}^a,t)}\biggl(-M g^{00}(q_{\cal C}^a,t)\frac{|\dot{q}_{\cal C}^a-\dot{\xi}^a|_{\xi^a=q_{\cal C}^a}|^2}{2}-\frac{1}{2}\sum_{{\cal C}^\prime (\neq {\cal C})}\Phi(|q_{\cal C}^a-q_{{\cal C}^\prime}^a|)\biggr)\\
&=&\sum_{\cal C}N_{\cal C}\biggl\{M\frac{1}{N_{\cal C}^2}\biggl(\frac{|\dot{q}_{\cal C}^a|^2}{2}-\dot{q}^a_{\cal C}N_{{\cal C},a}+\frac{|N^a_{\cal C}|^2}{2}\biggr)-\frac{1}{2}\sum_{{\cal C}^\prime (\neq {\cal C})}\Phi(|q_{\cal C}^a-q_{{\cal C}^\prime}^a|)\biggr\}\;,\label{eq:Lagfin2}
\end{eqnarray}
where we define
\begin{equation}
\Phi(|q_{\cal C}^a-q_{{\cal C}^\prime}^a|)=-\frac{M^2}{|{q}_{\cal C}^a- {q}^a_{{\cal C}^\prime}|}\;.\label{eq:potfin}
\end{equation}
Note that, for the space-time metric tensor $g_{\mu\nu}$ in the $3+1$ decomposition of the flat space-time (\ref{eq:31}), $\sqrt{-g}=N$ and $g^{00}=-1/N^2$ hold.
From the Lagrangian (\ref{eq:Lagfin2}), we obtain the canonical momentum (\ref{eq:Pdef}) of a condensate and the Hamiltonian (\ref{eq:Ham}) of the system in the vacuum state of quasi-particles.

\end{appendix}


\end{document}